\documentclass[%
 aip,
 jmp,%
 amsmath,amssymb,
 reprint,%
]{revtex4-1}

\usepackage{graphicx}
\usepackage{dcolumn}
\usepackage{bm}
\usepackage{mathtools}
\usepackage{hyperref}
\usepackage{mathrsfs}

\hypersetup{
    colorlinks,
    citecolor=black,
    filecolor=black,
    linkcolor=black,
    urlcolor=black
}

\begin{document}


\title[The Concept of Spin Ice Graphs and Field Theory for their Topological Monopoles and Charges]{The Concept of Spin Ice Graphs and Field Theory for their Topological Monopoles and Charges}

\author{Cristiano Nisoli}

\affiliation{Theoretical Division, Los Alamos National Laboratory, Los Alamos, NM, 87545, USA
}%


\date{\today}

\begin{abstract}
Spin ices can now be fabricated in a variety of geometries which control their collective behavior and exotic properties. Therefore,  their proper framework is graph theory. We relate spin ice notions such as ice rule, ice manifold, Coulomb phases, charges and monopoles, to graph-theoretical notions, such as balance, in/out-degrees, and Eulerianicity. We then propose a field-theoretical treatment in which topological charges and monopoles are the degrees of freedom while the binary spins are subsumed into entropic interaction among charges. We show that for a spin ice on a graph in a Gaussian approximation, the kernel of the entropic interaction is the inverse of the graph Laplacian, and we compute screening functions from the graph spectra as Green operators for the screened Poisson problem on a graph. We then apply the treatment on star graphs, tournaments, cycles, and regular spin ice in different dimensions.  

\end{abstract}

\maketitle


\section{Introduction}

Since the Bernal-Fowler ice rule~\cite{bernal1933theory} was invoked by Pauling~\cite{Pauling1935} to explain the zero-point entropy of water ice~\cite{giauque1933molecular,giauque1936entropy} the concept has come to describe a variety of other materials, such as pyroclore rare-earth spin ice antiferromagnets~\cite{Ramirez1999,den2000dipolar,Bramwell2001}, artificial magnetic spin ice antiferromagnets~\cite{tanaka2006magnetic,Wang2006,Nisoli2013colloquium,heyderman2013artificial,skjaervo2019advances}, or artificial particle-based ices~\cite{ortiz2019colloquium}.
Often~\cite{Ramirez1999,den2000dipolar,Bramwell2001,tanaka2006magnetic,qi2008direct,Morrison2013,nisoli2018frustration,gilbert2014emergent,Moller2009,Chern2011,Rougemaille2011}, but not always~\cite{Wang2006,Porro2013,zhang2013crystallites},  frustration impedes ordering in these materials, and leads to degenerate states of  constrained disorder, or {\it ice manifolds}, of interesting topological properties. 

Indeed, the ice rule is a topological concept related to the local minimization of a topological charge~\cite{nisoli2014dumping,nisoli2018frustration}. As such, it has a wide applicability, and artificial spin ice materials are being designed~\cite{nisoli2017deliberate} for a variety of emergent behaviors not necessarily found in natural magnets~\cite{nisoli2017deliberate,nisoli2018frustration,Zhang2013,gilbert2016emergent,gilbert2014emergent,lao2018classical,gliga2017emergent,farhan2017nanoscale,saccone2019dipolar}. Violations of the ice manifold are topological excitations~\cite{ryzhkin2005magnetic,Castelnovo2008} that, depending on the geometry and local degeneracy of the system, can be deconfined. Further, in magnetic materials these topological charges are also {\it magnetic charges}, and if deconfinable, {\em magnetic monopoles}~\cite{ryzhkin2005magnetic,Castelnovo2008,morris2009dirac,Giblin2011, ladak2011direct,Mengotti2010}. As such, they interact via a Coulomb law~\cite{Castelnovo2008}, are sources and sinks of the $\vec H$ field,  can  pin superconductive vortices in spin ice/superconductors heterostructures~\cite{wang2018switchable}, and could perhaps exert  a localized {\it and} mobile magnetic proximity effect~\cite{scharf2017magnetic} in heterostructures that interface two-dimensional spin ices  to transition metal dichalcogenides  or Dirac materials. Finally, in artificial realizations, these topological objects can also be read and written~\cite{wang2016rewritable,gartside2018realization}, and might therefore function as binary, mobile information carriers for spintronics for  for neuromorphic  computation~\cite{caravelli2018computation,arava2018computational}. 
 
In non-magnetic spin ices---such as particle-based ones~\cite{ortiz2019colloquium}, which can be made of confined colloids~\cite{Libal2006,ortiz2016engineering,loehr2016defect,libal2018ice}, superconducting vortices~\cite{libal2009,Latimer2013,Trastoy2013freezing,wang2018switchable}, skyrmions in magnets~\cite{ma2016emergent} or in liquid crystals~\cite{duzgun2019artificial,duzgun2019commensurate}---the mutual interaction among monopoles differs from a Coulomb law~\cite{libal2017,nisoli2018unexpected}. And yet, because they are  topological charges, they  always interact at least entropically~\cite{libal2018ice}, as we shall see.

In this work, first we extend the concept of spin ice on general graphs, and relate the two fields and their jargons, because many concepts of spin ice physics have been investigated under different names and with different aims in graph theory.

 Then we propose here a unifying framework for {\em degenerate} spin ice on a general graph. We treat these charge excitations as degrees of freedom, subsuming the underlying spin vacuum into  entropic interactions among charges.
This work has many motivations.

Firstly, a graph approach naturally separates geometry from topology. 
Graphs  possess a metric~\cite{goddard2011distance} yet are not necessarily embedded in a linear algebraic structure, each of them essentially describing a topological class of various and different geometric realizations. Results obtained on a graph are ipso facto related only to the topology of the system, rather than to geometry, thus gathering  under a unifying treatment  many topological concepts related to spin ice materials, some of which disseminated through many  works pertaining to a variety of specific systems~\cite{fradkin2013field,henley2010coulomb,castelnovo2010coulomb,henley2010coulomb,isakov2004dipolar,garanin1999classical,henley2005power,henley2011classical,ryzhkin2005magnetic,Castelnovo2008,castelnovo2012spin,ryzhkin2005magnetic,Giblin2011,castelnovo2011debye,kaiser2018emergent,castelnovo2010thermal,mol2009magnetic,farhan2019emergent,jaubert2011magnetic,Nascimento2012,mol2010conditions,dusad2019magnetic,klyuev2017statistics,kirschner2018proposal,castelnovo2010coulomb}. 

Secondly,  we also aim at  guiding future artificial realizations~\cite{Nisoli2013colloquium,heyderman2013artificial,skjaervo2019advances,ortiz2019colloquium,nisoli2018frustration} of exotic topologies~\cite{Morrison2013,Chern2013,nisoli2018frustration,gilbert2014emergent, gilbert2016emergent,lao2018classical,nisoli2018frustration}, including finite size systems~\cite{Li2010}.  For instance, interesting works on Penrose spin ices, based on finite size quasicrystals~\cite{bhat2013controlled,bhat2014non,brajuskovic2018observation,bhat2014ferromagnetic} still lack a proper language.

Thirdly, spin ice can now be realized in the quantum dots of a quantum annealer~\cite{king2020quantum} on a wide variety of graphs.

Finally, many common notions in spin ice physics posses a direct equivalent in graph theory. For instance: spin ice charges are the degree-excess between {\em indegrees} and {\em outdegrees};  the degeneracy of spin ice on a complete graph is simply the number of its {\em regular tournaments}; an ice rule configuration realizes a graph that can be unraveled in an Eulerian path, etcetera. A wealth of work has been done in graph theory, some of which could be easily translated to spin ice physics. This work represent an invitation extended both to graph theorist, to contribute to the field of spin ice physics, and to physicists working in spin ice, to broaden their mathematical approaches to these systems.

\section{ Spin Ice Graphs }

\subsection{Spin Ice}

Ice comes in about eighteen crystalline forms~\cite{petrenko1999physics},  all of them involving oxygen atoms residing at the center of tetrahedra, sharing four hydrogen atoms with four nearest neighbor  oxygen atoms (Fig.~1). Two of such hydrogens are covalently bonded to the oxygen of their molecule, and two realize  hydrogen bonds with oxygens of different molecules. Thus, two are ``in", two are ``out"  of the tetrahedron, and this is the so called {\em ice-rule}  introduced by Bernal and Fowler~\cite{bernal1933theory}. Each tetrahedron therefore realizes $6$ admissible configurations out of the $2^4=16$ ideally possible, and the collective degeneracy of the ice grows exponentially in the number of tetrahedra $N$ as $W^N$. This leads to a non-zero {\em residual} entropy per tetrahedron $k_B \ln W$.  Pauling famously estimated $W=3/2$, remarkably close to both the experimental~\cite{giauque1933molecular,giauque1936entropy} and  the numerically obtained value ($W=1.50685 \pm 0.00015$~\cite{nagle1966lattice}).

One can associate to the ice rule a spin model (Fig~1c) where binary spins are assigned to the bonds between molecules pointing toward the proton. Then the ice rule dictates that two spins point in, two point out, as is the case of magnetic {\em spin ices}. In rare earth titanates such as Ho$_2$Ti$_2$O$_7$ and Dy$_2$Ti$_2$O$_7$,  the magnetic cations Ho$^3+$ and Dy$^3+$ carry a very large magnetic moment, $\mu \sim10 \mu_B$. At low temperature they can be considered binary, classical Ising spins constrained to point along the directions of the lattice bonds which form a  pyrochlore lattice, and are thus expected to interact as magnetic dipoles obeying the ice rule~\cite{harris1997geometrical,Ramirez1999}.
\begin{figure}[t!]
\includegraphics[width=.9\columnwidth]{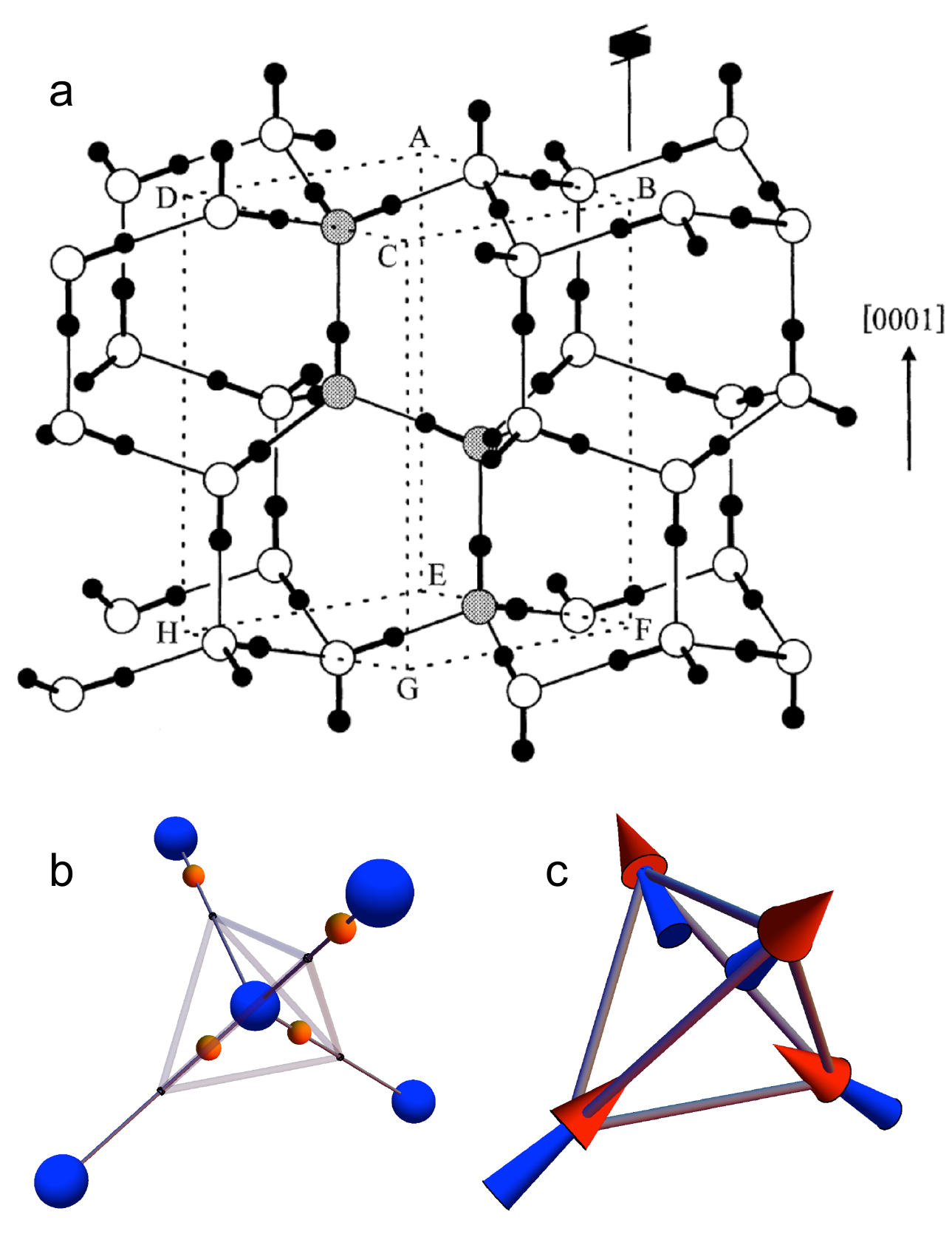}
\begin{center}
\caption{ a) The crystalline structure of ice shows proton disorder in the allocation of hydrogen atoms among  oxygen centered tetrahedra (image from ref~\cite{petrenko1999physics}). b) In water ice oxygen atoms sit at the center of tetrahedra, connected to each other by a hydrogen  atom. Two of such protons are close (covalently bonded) to the oxygen at the center, two are further away. c) One might replace this picture with spins pointing in or out depending on whether the proton is close or far away. Then two spins point in, two point out. This corresponds to the disposition of magnetic moments on pyrochlore spin ices, rare earth titanates whose magnetic ensemble does not order at low temperature, because of frustration, and, much like water ice, provides non-zero low temperature entropy density. (Figures b, c from ref~\cite{Castelnovo2008}).}
\label{default}
\end{center}
\end{figure}

While this {\em ice manifold} represents an interesting manifestation of a {\em Coulomb phase} which has been amply studied~\cite{henley2010coulomb,isakov2004dipolar,garanin1999classical,henley2005power} and which can be considered a prototype for {\em classical topological order}~\cite{henley2011classical,lamberty2013classical,castelnovo2012spin}, the exotic behaviors of spin ices proceed  not so much from said topological structure, but rather from how it is {\it broken}, e.g.\ via fractionalization into monopoles~\cite{ryzhkin2005magnetic,Castelnovo2008,castelnovo2012spin}, and how much of it is instead retained,  e.g.\ via spin fragmentation~\cite{brooks2014magnetic,canals2016fragmentation,petit2016observation}.

\subsection{Spins on a Graph}

We consider the most general case of a spin ice on a connected, undirected, simple graph~\cite{west2001introduction} $G$. The {\it  phase space} for $G$ is  the set of all {\em directed} graphs, o digraphs, that can be built on $G$ by assigning an orientation to its edges.

Consider an undirected, simple graph $G$ of a number $N_l$ of edges labeled by $l$, connecting a number $N_v$ of vertices labeled by $v$ and of various degree of coordination $z_v$. We call $\{ vv'\}$ an edge $l$ among vertices $v, v'$. For such graph, the {\em adjacency matrix}~\cite{west2001introduction} is the matrix  $A_{vv'}$, such that  $A_{vv'}=1$ if $v, v'$ are connected and $A_{vv'}=0$ otherwise. It contains all the information of the graph. Obviously, $A_{vv'}$ is symmetric and $z_v=\sum_{v'} A_{vv'}$.

We can define binary variables or Ising spins $S_l$ on each edge $l$, as in Fig.~2, via an antisymmetric matrix $S_{vv'}$ such that $S_{vv'}=0$ if $v$ and $v'$ {\it do not} share an edge, $S_{vv'}=1$ if they do and the spin points toward $v'$, and $S_{vv'}=-1$  if they do and the spin points toward $v$.  

In the language of graph theory, $A_{vv'}$ defines an undirected graph, while an Ising spin structure $S_{vv'}$ defines a  digraph. 
The $S$ matrix  is related to the non-symmetric adjacency matrix of the corresponding digraph, $A^{\mathrm{dir}}_{vv'}$, whose elements have value one if and only if $vv'$ are connected by an edge pointing toward $v'$ (on simple graphs). Then $S$ is the anti-symmetrization of $A^{\mathrm{dir}}$, or $S_{vv'}=A^{\mathrm{dir}}_{vv'}-A^{\mathrm{dir}}_{v'v}$ (and of course  $A_{vv'}= A^{\mathrm{dir}}_{vv'}+A^{\mathrm{dir}}_{v'v}$).

For an undirected graph $G$, we call its {\it  phase space}  ${\cal P}$,  the set of  the  digraphs that can be specified on it. Clearly, the cardinality of the phase space is $|{\cal P}|=2^{N_l}$. 

\subsubsection{Ice Manifolds}

An {\it ice manifold} is a proper subset of the phase space ${\cal P}$ that minimizes the topological charge, defined as follows. 

Given $S_{vv'}$, its topological charge distribution is the vector $Q_v$  defined for each vertex $v$ as the difference between the edges pointing in and out of $v$, or
\begin{equation}
Q_v[S]=\sum_{v' } S_{v'v}.
\label{charge}
\end{equation}
In graph theory, $Q_v$ is thus the difference between {\it indegrees} and {\it outdegrees} of the directed graph that corresponds to a particular spin configuration on $G$. $Q_v$ can have the values $z_v, z_v-2, \dots, 2-z_v, -z_v$, and thus only vertices of even coordination can have zero charge. 
To relate  Eq.~(\ref{charge}) to the more physical picture~\cite{henley2010coulomb}, we can introduce the divergence operator on a graph as
\begin{equation}
\mathrm{div}[S]_v=\sum_{vv'}S_{vv'}
\label{div}
\end{equation}
from which we immediately have
\begin{equation}
Q_v[S]=-\mathrm{div}[S]_v.
\label{charge2}
\end{equation}
\begin{figure}[t!]
\includegraphics[width=.999\columnwidth]{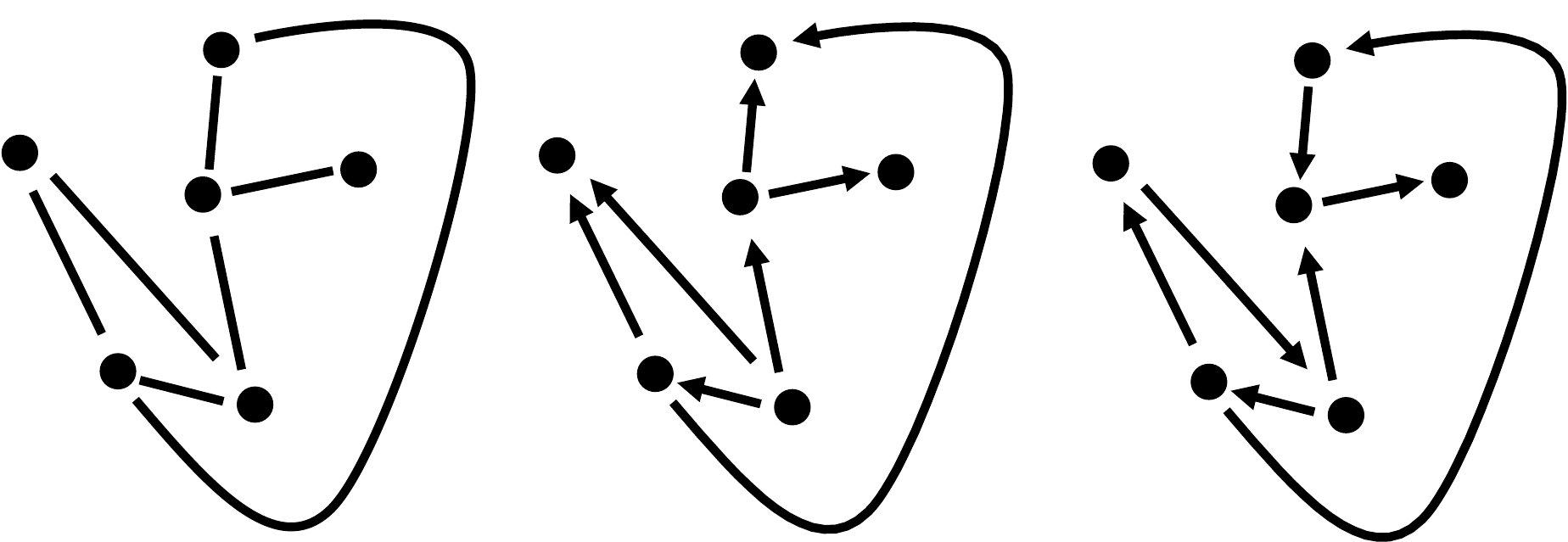}
\caption{Left: undirected graph. Center: directed graph. Right: directed graph obeying the ice rule.  }
\label{fig1_0}
\end{figure}

A digraph $S_{vv'}$ (i.e.\ a spin configuration) for which $|Q_v|$ is minimal on each vertex, i.e.\ has zero charge on all vertices of even coordination and $\pm1$ charge on all vertices of odd coordination, is said to {\it obey the ice-rule}.  

Then, given a graph $G$, we call the {\it ice manifold} of $G$ the subsets ${\cal I} \subset {\cal P}$ of its spin phase space ${\cal P}$  made of all directed graphs that obey the ice rule. Then, $|{\cal I}|$ is the ice-degeneracy of the graph and $s=N_l^{-1}\ln|{\cal I}|$ is its {\em Pauling entropy} per spin. 

In the language of graph theory, an ice rule configuration corresponds to a {\em balanced} or {\em quasi-balanced} directed graph. By a famous theorem, if all vertices have even degrees (i.e.\ even coordination) a balanced digraph is  {\em Eulerian}. That means that an Eulerian path exists (a walker can follow the arrows and walk the entire graph without passing the same edge twice~\cite{euler1736commentarii,biggs1986graph}). Moreover, if the graph is complete and has odd vertices, a spin configuration is called a {\em tournament} because it describes winners and losers in a round-robin tournament~\cite{landau1953dominance,harary1966theory}. Thus, an ice rule configuration on a complete graph describes a {\em regular} tournament, one where all players score the same. Therefore, the ice-degeneracy of a complete graph of odd number of vertices counts all the possible outcomes of a round-robin tournament where every player beats the same number of other players (which is still an open problem).  
 
We see therefore that spin ice physics is  a case of the theory of balanced or Eulerian graphs. The latter is greatly consequential, with applications in biology~\cite{pevzner2001eulerian}, computer science~\cite{khodjasteh2009dynamically}, logistics~\cite{bayen2006adjoint}, and sociology~\cite{roberts1978graph}. In complex networks, digraphs can describe the sharing of a token among agents, and deviation from the ice-rule can quantify the agent satisfaction or frustration, and give measures of fairness~\cite{mahault2017emergent}.

Depending on the topology of the graph it is not obvious  if and when the cardinality of the ice manifold scales exponentially with the size of the graph, thus leading to a non-zero Pauling entropy~\cite{Pauling1935}. Indeed, the one dimensional ferromagnetic Ising model can be mapped into a spin ice on a path graph, or a cycle, and its ice manifold has cardinality two, regardless of the number of vertices. We will show elsewhere that the path graph is the only such case.  

\subsubsection{Coulomb Phases}

The concept of a Coulomb phase appeared first in gauge field theories~\cite{fradkin2013field} and was then introduced in the theory of  pyrochlore spin ice~\cite{isakov2004dipolar,henley2005power, henley2010coulomb}. In simple terms, it corresponds to a disordered spin texture that can be coarse-grained to a solenoidal magnetization field. It can be considered a case of classical topological order~\cite{henley2011classical,lamberty2013classical,castelnovo2012spin} where no order parameter exists but instead the disordered states are labeled by gauge fields whose solenoidal nature expresses the constraint over the disorder. Excitations are then violations of the solenoidal requirement. 

We can generalize this notion to a graph spin ice by expressing it without coarse graining or gauge fields. 
We say that two spin assignations  $S, S' \in {\cal P}$ are {\it charge equivalent} if and only if the difference $(S_{vv'}-S'_{vv'})/2$ has {\it zero} topological charge on every vertex. 
Charge-equivalence is an equivalence relation and thus induces a partition on the phase space ${\cal P}$. We call each class of equivalence in that partition a {\em Coulomb class}. Each Coulomb class is labeled by a distribution of charge $Q_v$.

A trivial example: a graph made of only two  vertices connected by one edge has a spin phase space of cardinality $2$, corresponding to the two orientations of its only spin. Its ice manifold coincides with the spin phase space, and it contains $2$ Coulomb classes, each of cardinality $1$. 
For a less trivial example, the ice manifold of pyrochlore ice is a Coulomb class corresponding to charge $Q_v=0$ everywhere. The same is true by definition for every graph that has only   even coordinations. 

Clearly, not all ice manifolds are Coulomb classes. For instance the ice manifold of the honeycomb graph, so-called kagome ice, is not: for any spin configuration there is always at least one spin, and in fact an extensive number of spins, that can be flipped individually, thus changing charge configurations (and Coulomb class) without violating the ice rule. And yet, the kagome ice manifold can be partitioned into Coulomb classes. Crucially,  its  Ice II phase~\cite{Moller2009,Chern2011,Rougemaille2011,chioar2016ground,libal2017} corresponds to two Coulomb classes, each of charge alternating in sign $\pm1$ on nearby vertices. Indeed, the spontaneous symmetry breaking between the two Coulomb classes drives a second order transition to charge ordering (of the Ising universality class).

A Coulomb class  becomes important if it represents a low energy state. In such case it defines a {\em Coulomb phase}.  Thus, the ice manifold of  spin ices of  even coordination, including therefore pyrochlores,  is a Coulomb class and also a Coulomb phase. In kagome ice,  each of the two Coulomb classes of the Ice-II phase are  Coulomb phases, which might explain why it is so hard to observe it experimentally~\cite{Rougemaille2011,macdonald2011classical,Zhang2013,drisko2015fepd}, whereas the Ice-I phase is found easily~\cite{tanaka2006magnetic,qi2008direct,Nisoli2010}. 

Indeed,  a Coulomb class imposes topological constraints on the kinetics within the class: it  prohibits single spin flips and requires collaborative flips (corresponding to Euler trails)  that might be extremely unlikely in a realistic dynamics. Thus, in a Coulomb phase, all kinetics must happen {\em above} that phase, by breaking the topology of the Coulomb class, that is by changing its defining distribution of charges. In pyrochlore ices or in degenerate artificial square ices, where the entire ice manifold is a Coulomb phase, the breakage of topological protection consists in the appearance of monopoles, that is $\pm 2$ charges. In  such systems all single spin kinetics is then monopole kinetics. In the Ice-II phase of kagome ice, also a Coulomb phase as noted by McDonald {\em et al.}~\cite{macdonald2011classical}, kinetics consists in the  breaking of the $\pm 1$ charge alternation. 

Because  properties of topological order pertain to  Coulomb classes, which are defined by charges, we  develop a formalism in which charges are the relevant degrees of freedom. We  reach an effective free energy for charge distribution, which contains the entropic effect of the corresponding Coulomb class.

\section{Field Theory of Emergent Charges}

We need a Hamiltonian whose ground state is the ice manifold, and thus allows us to deal with breakages of the topological protection in terms of  energy costs. 
The following is minimal
\begin{equation}
{\cal H}[S] = \frac{\epsilon}{2}\sum_v Q_v^2, 
\label{Hs}
\end{equation}
where $\epsilon>0$ is an energy. Note that ${\cal H}$ more generally {\em measures} how ``unbalanced'' a digraph is. Note also that for every digraph $\sum_v Q_v^2$ is larger than or equal to the number of vertices with odd coordination.

The corresponding partition function reads
\begin{equation}
Z[H, V]=\sum_{\{S_{vv'}\}} e^{-\beta \left({\cal H}[Q]- \frac{1}{2}\sum_{vv'}  S_{vv'} H_{vv'} -\sum_v V_v Q_v\right)},
\label{Z}
\end{equation}
where $\beta=1/T$, $H_{vv'}$ is an antisymmetric matrix modeling an external ``magnetic'' field defined on the edges with respect to the orientation $l=vv'$, so that $H_{vv'}\ne0$ if and only if $v$ and $v'$ share an edge, and $H_{vv'}=-H_{v'v}$. $H$ has dimensions of an energy (the Zeeman energy), and $V_v$ is an external potential acting on the charges.

Clearly then
\begin{align}
\langle S_l\rangle &=T\frac{\delta \ln Z[H,V]}{\delta {H_l}} \nonumber \\
 \langle S_l S_{l'}\rangle &=T^2\frac{\delta^2 \ln Z[H,V]}{\delta {H_l}\delta 
{H_{l'}}} \nonumber \\  
\langle Q_v\rangle &=T\frac{\delta \ln Z[H, V]}{\delta {V_v}}\nonumber \\  
\langle Q_vQ_{v'}\rangle &=T^2\frac{\delta^2 \ln Z[H, V]}{\delta {V_v}\delta {V_{v'}}}
\label{Sav}
\end{align}
are the one- and two-point correlation functions for  spins and charges. 
%



To produce a field theory we need to remove the discrete variables $S_l$. We do so by a common trick. We  insert in the sum of (\ref{Z}) the tautological  expression {$1=(2\pi)^{-N_v}\prod_v \int   dq_v d\phi_v\exp\left[i \phi_v\left(q_v-Q_v\right) \right]$} and then we sum over the spins. We introduce 
\begin{align}
\Omega[\phi, H]&=\sum_{\{S_{vv'} \}}e^{-i\sum_v\phi_vQ_v+\frac{\beta}{2}\sum_{vv'}  S_{vv'} H_{vv'}} \\
&={\prod_{{\langle vv' \rangle}}}\sum_{S=\pm1}e^{[-i(\phi_{v'}-\phi_{v})+\beta H_{vv'} ]S}
\end{align}
where ${\langle vv' \rangle}$  are edges (counted once, so that if ${\langle vv' \rangle}$ is counted ${\langle v'v \rangle}$ is not). From it we obtain 
\begin{equation}
\Omega[\phi,H]=2^{N_l}{\prod_{{\langle vv' \rangle}}}\cos\left(\nabla_{vv'}\phi + i \beta H_{vv' } \right),
\label{F}
\end{equation}
 where the matrix $\nabla \phi$ is defined on the edges only as
\begin{equation}
\nabla_{vv'}\phi = \phi_{v'} - \phi_{v}
\label{discgrad}
\end{equation}
and is called the {\it gradient} matrix of $\phi$. Note that if the graph is embedded in a linear space and $\vec{vv'}$ is the  vector pointing from $v$ to $v'$ then $\nabla_{vv'}\phi=\vec{vv'} \cdot \vec \nabla\phi + O(|vv'|^2)$.

 $Z[H,V]$ from (\ref{Z}) can now be rewritten as
\begin{equation}
Z[H,V]=\int \left[ dq \right] \tilde\Omega[q,H] e^{-\beta{\cal H}[q]+ \beta\sum_v  q_v V_v}
\label{Z2}
\end{equation}
%
%
where  $\left[ dq \right]= \prod_v dq_v $, ${\cal H}[q]$ is given by Eq.~(\ref{Hs}), and $\tilde \Omega[q,H]$  is the functional Fourier transform of $\Omega[\phi,H]$, or 
\begin{equation}
\tilde \Omega[q,H]=\frac{1}{(2\pi)^{N_v}}\int [d\phi] \Omega[\phi,H] e^{\sum_v i q_v \phi_v}.
\label{F}
\end{equation}
The averages and correlations of $q$ are the same as for $Q$ and therefore, for instance $\langle q_v\rangle= \langle Q_v\rangle$ and also
\begin{align}
\langle \mathrm{div}[S]_v \mathrm{div}[S]_{v'}\rangle=\langle q_v q_{v'}\rangle.
\end{align}

We have thus replaced the binary spin variables $S_l$ defined on the edges with a continuum field $q_v$ defined on the discrete set of vertices,  but we have  gained a term $\tilde \Omega[q]$. It represents a generalized degeneracy or density of states for the charge distribution $q_v$, emergent from the many possible underlying spin ensembles compatible with $q_v$. It also constraints $q_v$ to proper, discretized values. 

We can call ${\cal S}=\ln \tilde \Omega[q]$ the generalized entropy for the charge distribution $q$. Then,  the effective free energy at zero loop~\cite{zinn1996quantum}  for $q_v$ is given by the quadratic part of ${\cal H}[q]-T{\cal S}[q]$ as we shall see. Also, in absence of a field, from (\ref{F}), (\ref{Sav}) and the parity of the functions involved, $\langle q_v\rangle= 0$  for every $v$  and $\langle S_l\rangle=0$ for every $l$, as one would indeed expect from trivial considerations on the model.

To see the same formalism from a different angle we can rewrite the partition function as
\begin{equation}
Z[H,V]=\frac{1}{(2\pi)^{N_v}} \int \left[ dq d\phi \right] e^{-\beta{\cal F}[q,\phi,H]}  
\label{QFT}
\end{equation}
with
\begin{equation}
{{\cal F}[q,\phi,H]} ={\cal H}[q]-\sum_v q_v \left(  iT \phi_v +V_v \right)+ {\cal F}[\phi, H]
\label{QFT2}
\end{equation}
and
\begin{equation}
{\cal F}[\phi, H]=-T \ln \Omega[\phi,H].
\label{QFT3}
\end{equation}
Equations (\ref{QFT}), (\ref{QFT2}) look familiar in the language of quantum field theory. They correspond to a charge  field $q_v$, for which $i T \phi_v$ acts  as a bosonic field (of ``Lagrangian''  ${\cal F}[\phi]$) mediating an interaction   between charges. Again, the interaction is not real but comes from the underlying binary ensemble from which the charge field is an emergent observable. 
Both pictures come useful in different scenarios, as we shall see. 

Integrating  (\ref{QFT}), over $dq_v$ we obtain 
\begin{equation}
Z[H, V]=\left(\frac{T}{2\pi \epsilon}\right)^{\frac{N_v}{2}}\int \left[ d\phi \right]  \Omega[\phi,H] e^{-\frac{T}{2\epsilon} \sum_v \left( \phi_v -i \beta V_v\right)^2} 
\label{Z3}
\end{equation}
which shows that in general $\langle i \phi \rangle$ is real. Indeed from Eqs.~(\ref{Sav},\ref{Z3}) we have
\begin{equation}
\epsilon \langle q_v \rangle = i T\langle \phi_v \rangle  +V_v.
\label{gigi}
\end{equation}

To further clarify the intuitive meaning of $i\phi$ as an emergent, entropic field translating the effect of the spin correlations, from  Eq.~(\ref{Z3}) we can immediately deduce 
  \begin{equation}
 \langle S_{vv'}\rangle =\langle \tanh \left(\beta H_{vv'}+i\nabla_{vv'}\phi \right) \rangle .
 \label{M}
 \end{equation}
It is then natural to introduce an entropic ``magnetic field'' $B^e_{vv'}=i \nabla_{vv'}\phi$, and then
\begin{equation}
\langle S_{vv'}\rangle  \simeq  \beta H_{vv'} +   \langle  B^e_{vv'} \rangle 
\label{M2}
\end{equation}
when $ \langle S_{vv'}\rangle$ is small (which requires $H_{vv'}$ small). Note than in general, from Eq.~(\ref{gigi}), the entropic magnetic field is related to the gradient of the charges, or
\begin{equation}
\epsilon \nabla_{vv'} \langle q \rangle = \langle B^e_{vv'} \rangle  +E_{vv'}
\label{gigi2}
\end{equation}
where $E$ is the gradient of $V$.

It is time to confess that $V_v$ is superabundant, though useful. Indeed, nothing changes by incorporating $V_v$ into $H_{vv'}$ by replacing in our equations $V \to 0$, $H_{vv'} \to H_{vv'}+\nabla_{vv'} V$. Similarly, if $H$ can be divided into a gradient plus a term irreducible to a gradient, or $H_{vv'}=H'_{vv'}+\nabla V'_{vv'}$, the equations above are still valid with the substitution $V\to V+ V'$, $H \to H'$.  In the following we will assume $V=0$.

Finally, taking the divergence of Eq.~(\ref{M}), we obtain 
  \begin{equation}
 \langle q\rangle = - \text{div}\left[\langle \tanh \left(\beta H+i\nabla \phi \right) \rangle \right].
 \label{self}
 \end{equation}
In the linearized limit in which Eq.~(\ref{M2}) is valid, we obtain
  \begin{equation}
 \langle q\rangle = - \beta \mu~\! q_{\text{ext}} + i \hat L \phi.
 \label{self2}
 \end{equation}
In the previous equation the {\em external charge} is defined as $\mu q_{\text{ext}}=\text{div}[H]$, where $\mu$ is an energy, and the {\em Laplace operator} $\hat L$ is defined via the {\em Laplacian matrix}
\begin{equation}
L_{vv'}=z_v \delta_{vv'}-A_{vv'}, 
\label{D2}
\end{equation}
or $\hat L=\hat D-\hat A$ where $D_{vv'}=z_v \delta_{vv'}$ is the  {\em degree matrix} and $z_v$ is the degree or coordination of the vertex $v$. The Laplacian matrix is  the generalization on a  graph of the discretized Laplacian operator on a lattice. The reader can easily verify that  on a square lattice of edge length $a$ when one takes the usual continuum limit for $a\to 0$ one finds $\hat L\to -a^2\nabla^2$. Also, the reader can verify that for a generic  $\zeta_v$ defined on the nodes $v$
\begin{equation}
\mathrm{div}[\nabla \zeta]_v=-\sum_{v'} L_{vv'}\zeta_{v'}
\end{equation}
as one would expect as the generalization of the notorious $\vec \nabla \cdot \vec \nabla=\nabla^2$, valid in linear spaces, and which we have used to deduce Eq.~(\ref{self2}). 

Thus, Eq.~(\ref{self2})  tells us that charges are the sources and sinks  of the entropic potential, and represent a generalization of the at least in a linear approximation. There, the entropic field $i\phi$ obeys a generalized Poisson equation and is therefore the {\em Coulomb potential} of the charges on a graph. We shall see in the last section that indeed when the graph can be properly embedded in a linear space, the entropic interaction among charges is indeed the standard Coulomb interaction in the proper dimension of the space.

\section{High $T$ Approximation}

%
 
By taking the high $T$, high $H$ limit  but keeping $H/T$  finite, the Gaussians in (\ref{Z3}) tend to delta functions in $\phi$ and we obtain   
\begin{equation}
 Z[H]=2^{N_l}\prod_l \cosh\left( \beta H_l\right). 
 \label{para}
\end{equation}
Unsurprisingly, the above is the standard ``paramagnetic'' partition function for an uncorrelated system. It leads, via Eqs.~(\ref{Sav})  to the familiar magnetization law for a paramagnet
 \begin{equation}
 \langle S_l\rangle=\tanh(\beta H_l).
 \label{trivialM}
 \end{equation}
 When  $H=0$, we get  from (\ref{para}) the correct entropy per spin at high temperature, or  $s=\ln2$. Thus under assumption of disorder, the high $T$ expansion corresponds to an expansion in the small entropic field $\phi$. In doing so we lose the constraint to discrete values of $q_v$ imposed  by $\Omega[\phi]$, which is however not relevant at high $T$.


\subsection{Free Energy and Entropic Interactions}

 We can  expand Eq.~(\ref{QFT3}) at lowest order and find
\begin{align}
\beta{\cal F}_H[\phi] &=-\frac{1}{2} \sum_{vv'}  A_{vv'} \ln \cos \left(\nabla_{vv'}\phi +i\beta H_{vv'}\right) \nonumber \\
 &\simeq \frac{1}{4 }\sum_{vv'} A_{vv'} \left( \nabla_{vv'} \phi + i\beta H_{vv'} \right)^2 \nonumber \\
 &= \frac{1}{2}\sum_{vv'} \phi_v \left(z_v \delta_{vv'}-A_{vv'} \right) \phi_{v'}  \nonumber \\
& - i\beta \sum_{vv'} \phi_v A_{vv'} H_{vv'}  -\frac{\beta^2}{4} \sum_{vv'} A_{vv'} H_{vv'}^2,
\label{QFT4}
\end{align}
and thus write ${\cal F}$ in Eq.~(\ref{QFT2}) at the second order  as
\begin{align}
\beta{\cal F}_2[q,\phi]&=\frac{\beta \epsilon}{2} q^2 + \frac{1}{2} \phi\hat L \phi - i \phi \cdot \left(q + \beta \mathrm{div}[\hat H]\right)-\frac{\beta^2}{2} |H|^2
\label{QFT5}
\end{align}
where matrices are expressed as operators acting on  $q$, $\phi$, which are vectors of dimension $N_v$, $q^2=q\cdot q=\sum_vq_v^2$, and  $|H|^2=\sum_{vv'} A_{vv'} H_{vv'}^2/2$. 

Note that Eq.~(\ref{QFT5}) is correct {\it only because} the divergence of $H$ is hereby  defined discretely with respect to vertices by Eq.~(\ref{div}). One might at first suspect that it is incorrect, and immagine that, e.g.\, on a square lattice a continuous solenoidal $H$ would not couple to the spin ice. However, a moment thought shows that even for an uniform $H$,  its divergence, as defined {\it on vertices} in Eq.~(\ref{div}),  is necessarily non-zero at the boundaries  of the lattice even for a uniform field. In other words, the definition of divergence on a graph already accounts for boundary conditions (inclusive of ``internal'' boundaries~\cite{libal2018ice}).

By integrating over $q_v$ we can express the partition function in Eq.~(\ref{QFT}) in the high $T$ approximation as coming from a free energy in the entropic field only, or
\begin{align}
\beta{\cal F}_2[\phi]&=\frac{1}{2}  \frac{T}{\epsilon} \phi^2+ \frac{1}{2} \phi \hat L \phi - i \phi \cdot \left(\beta \mathrm{div}[\hat H]\right)-\frac{\beta^2}{2} |H|^2.
\label{pippo}
\end{align}

In a similar way, to find an effective free energy for the charges,  using Eqs~(\ref{QFT}), (\ref{QFT5}), we can  integrate instead over the entropic fields $\phi$. However, to do so, we must first consider the spectrum~\cite{brouwer2011spectra} of $\hat L$. 

The following is well known from the spectral theory of graphs~\cite{}. $\hat L$ is symmetric and thus has real eigenvalues $\{\gamma(k)^2\}_{k=0\dots k_{\max}}$ with  $k_{\max} \le N_v-1$, and corresponding $N_v$  eigenvectors $\psi^{\alpha}_v(k)$ (where $\alpha$ counts the eigenvalue degeneracy). 
It is  immediate to verify that  $\gamma^2(0)=0$ for the  uniform eigenvector $\psi_v(0)=1/\sqrt{N_v}$. In a simple and connected graph all other eigenvalues are strictly positive. 

We can go to the new basis, defining $\tilde q^{\alpha}(k)={{\psi^{\alpha}}^*(k)\cdot q}$ and $\tilde \phi^{\alpha}(k)={{\psi^{\alpha}}^*(k)\cdot \phi}$. Then in  Eqs~(\ref{QFT}), (\ref{QFT5}), the integration over $d\tilde \phi(0)$ merely returns a  $\delta(\tilde q(0))$, which in ``real space'' corresponds to $\delta\left(\sum_v q_v\right)$. This ensures  that in the new free energy we sum only over charge configurations of zero net charge--as it should be, since a system of dipoles is charge neutral. All other charge modes have zero net charge. Indeed  for any eigenvector  of $L$ except the one of zero eigenvalue it is true that   $\sum_v \psi_v(k)=0$. This follows immediately from $\sum_v L_{vv'}=z_{v'}-z_{v'}=0$ and $\psi_v(k) =\sum_{v'} L_{vv'} \psi_{v'}(k)/\gamma(k)^{2}$. 

From Gaussian integration of ${\cal F}_2$ in Eq.~(\ref{QFT5}) in the space orthogonal to $\psi(0)$ (where $\hat L$ can be inverted)  we obtain, in absence of field $H$, the effective free energy for $q$  in the form
\begin{align}
\beta{\cal F}_2[q]&=\frac{\beta \epsilon}{2} q^2 + \frac{1}{2} q{\hat L}^{-1} q,
\label{QFT6}
\end{align}
which can be interpreted both in real or spectral space. 

The first term is the usual energy cost to produce charges.  The second term tells  us that,  at quadratic order, the effect of the underlying spin manifold   can be subsumed into a pairwise, entropic interaction that corresponds to $T L^{-1}$, i.e.\ the Green operator of the Laplacian.

In regular lattices embedded in a linear space (see below),  Eq.~(\ref{QFT6}) implies that in three dimensions (3D) charges interact entropically via a $1/x$ law, as indeed found numerically~\cite{}. In two dimensions (2D) one expects instead a logarithmic interaction. In both case we have Coulomb potentials in the proper dimension, as anticipated in our discussion of Eq.~(\ref{self2}).

This, however, also implies a mismatch in systems of reduced dimensionality, for instance in square ice, whose entropic interaction is the Green function of the 2D laplacian while the real monopole interaction (not considered in this work) is the Green function of the 3D laplacian. Such mismatch leads to lack of screening, as we show elsewhere~\cite{}.

\subsection{Charge Correlations}

We  define $\hat W^q$ as the inverse of the kernel of the free energy for the charge of Eq.~(\ref{QFT6}), or
\begin{align}
\hat{W^q}^{-1}=\hat 1 \beta \epsilon + \hat {L}^{-1},
\label{W}
\end{align}
and then, from equipartition we have
\begin{align}
 \langle q_v q_{v'}\rangle =W^q_{vv'} 
\label{qcorr5}
\end{align}
We note that $\hat W^q$ can be written in various ways, including
\begin{align}
\hat W^q = \hat L(\xi^2\hat L +1)^{-1} 
=\xi^{-2}- \xi^{-4}\hat G,
\label{W2}
\end{align}
where we call
\begin{equation}
\xi=\sqrt{\epsilon/T}
\label{xi0}
\end{equation}
 the {\it correlation length at high temperature}, as it was   already appreciated in more specific systems, via other means~\cite{garanin1999classical,henley2005power} (see next section).
In  Eq.~(\ref{W2}) we have introduced  the ``Green function'' (actually, a matrix) of the screened Poisson equation on a graph, or
\begin{equation}
\hat G= \left(\hat L +\xi^{-2}\right)^{-1},
\label{G}
\end{equation}
which, as we show in the next subsection, controls the screening from an external charge. Note that, from Eq.~(\ref{pippo}) $\hat G$ is also the correlation for the entropic field $\phi$, or
\begin{align}
 \langle \phi_v \phi_{v'}\rangle =G_{vv'} .
\label{phicorr}
\end{align}

Finally, going to the spectrum of $L$ we find  
\begin{align}
\langle {{\tilde q}^{\alpha}}{^*}(k) \tilde q^{\alpha'}(k') \rangle&=\delta_{\alpha \alpha'} \delta_{kk'} w(k) \nonumber \\
 &=\delta_{\alpha \alpha'} \delta_{kk'} \frac{\gamma(k)^2}{ \gamma(k)^2\xi^2+1}.
\label{qcorr1}
\end{align}
%
%
In the infinite temperature limit the correlation length becomes zero, and from Eq.~(\ref{qcorr1}) we obtain
\begin{equation}
\langle q_v q_{v'}\rangle \to L_{vv'} ~~\mathrm{for} ~~ T\to \infty.
\label{qcorr2}
\end{equation}
From Eq.~(\ref{qcorr2}) we have  
\begin{equation}
\langle q_v^2\rangle \to z_v~~~\text{for}~~~T\to \infty
\label{qsatur}
\end{equation}
 which  indeed corresponds to the average square charge of uncorrelated vertices, or $ \overline {q^2}_{\text{uncorr}}$ defined as the square charge obtained from counting arguments. Indeed, considering vertex multiplicities only, and computing the  average square charge for a vertex of degree $z$ with  each charge $2n-z$  weighted merely by its vertex multiplicity ${z} \choose {n}$, one obtains 
\begin{equation}
\overline{q^2}_{\text{uncorr}}=2^{-z}\sum_{n=0}^z (z-2n)^2 {{z} \choose {n}}=z.
\label{quncorr}
\end{equation}
Note also that Eq.~(\ref{qcorr2}) implies zero correlations among vertices that are not nearest neighbors, but a correlation of $-1$ among neighboring vertices: since nearby vertices share a spin, they have in average opposite charges.

How is the infinite temperature limit approached? From Eq. (\ref{qcorr1}) we  obtain, at least formally, the series
\begin{align}
\langle q_v q_{v'}\rangle= \left[L \sum_{n=0}^{\infty} \left(-\xi^2 L\right)^{n} \right]_{vv'}. 
\label{qcorr3}
\end{align}
%
Note now that $L^n$ can be written as sums of products of $n$ $\hat A$ and $\hat D$ matrices (e.g. $ADDAADAD\dots$). A moment's thought should convince that if $v$, $v'$ are separated by more edges than the number of $A$ matrices in such product, then the $vv'$ element of the product is zero. An obvious notion of  distance between two vertices on a graph is given by the number of edges in the shortest {\it path} (aka graph geodesic) connecting them~\cite{aouchiche2014distance}. It follows that if $v$ and $v'$ are at a distance $d_{vv'}>1$ the first nonzero term of the series in Eq.~(\ref{qcorr3}) is given by
\begin{align}
\langle q_v q_{v'}\rangle= \left(-\xi^2 \right)^{(d_{vv'}-1)} \left[A^{d_{vv'}}\right]_{vv'} +O\left(\xi^{2d_{vv'}}\right).
\label{qcorr4}
\end{align}
%
%
Interestingly, $A^k_{vv'}$ is known  to be the number of {\it walks} of length $k$ between the two vertices $v$ and $v'$. Thus, the coefficient $ \left[A^{d_{vv'}}\right]_{vv'} $ in Eq.~(\ref{qcorr4}) is the number of walks between the two vertices $v$ and $v'$ of length equal to the distance $d_{vv'}$. It is therefore the number of  geodesics connecting the two vertices. Note that this considerations are conditional to the possibility of the expansion of Eq.~(\ref{qcorr3}), which might not hold in the thermodynamic limit of certain systems. 

By construction our approximation doesn't work for $T\to0$ where the fluctuations of the entropic field diverge.  One could speculate that by including perturbative terms functional forms would not change, except for replacing $\xi(T) \to \xi_r(T)$ where $\xi_r(T)$ is the real correlation length at low $T$. For instance in pyrochlore ice, correlations become screened-algebraic at low $T$, suggesting that a quadratic free energy might work with proper renormalizations of the parameters.   

In the limit $\xi\to\infty$ From Eq.~(\ref{qcorr1}) we find 
\begin{align}
\langle q_v q_{v'}\rangle= \xi^{-2}\delta_{vv'}-\xi^{-4}\left[L^{-1}\right]_{vv'} +O(\xi^6).
\label{trippa}
\end{align}
Then, for graphs of even coordination we obtain
\begin{equation}
\xi^2 \sim 1/\langle q^2\rangle~~\text{for}~~ \xi^2 \to \infty,
\label{xir}
\end{equation}
 Note that the equation above corresponds to the Debye screening length for a Coulomb potential whose coupling constant is proportional to $T$, which is indeed the case of our entropic field. At high temperature, the situation is much different, with $\xi^2=\epsilon/T$. The Debye-H\"uckel approach applies to strong electrolytes that are fully dissociated, and the disorder brought by higher temperature prevents charges from screening, thus increasing the screening radius. In spin ice, instead, charges carry an energy cost  and at higher temperature there are more charges available for the screening.

Furthermore, Eq.~(\ref{xir})   corresponds in  pyrochlore spin ice  to the experimentally found~\cite{fennell2009magnetic} exponentially divergent behavior of the correlation length in the proximity of the ice manifold. And because for $v\ne v'$ the correlations tend to $\propto \hat L^{-1}$, we call them {\it Coulomb} in this limit.

\subsection{Entropic Screening of External Charges}

Since we have no  real interaction among charges, all screening is entropic.  One can consider two cases: screening of an external charge and screening of a pinned charge.

One can define external charges as sources and sinks of the $H$ field, which interact with emergent spin ice charges via the coupling of the $H$ field to the spins $S$. 
To understand this formally, consider the term $\sum_{vv'}S_{vv'}H_{vv'}$ in Eq.~(\ref{Z}). Imagine that we can write a Helmholtz decomposition on the graph, so that $H$ can be represented as $H_{vv'}=\nabla_{vv'}\Psi +H_{vv'}^{\perp}$ where $\Psi_v$ is a field and the second term has no divergence. Then we have
\begin{align}
\frac{1}{2}\sum_{vv'}S_{vv'}H_{vv'}=-\sum_vQ_v\Psi_v+ \frac{1}{2}\sum_{vv'}S_{vv'}H_{vv'}^{\perp},
\label{Hel}
\end{align}
that is, the potential responsible for the divergence-full part of the magnetic field couples to the emergent spin ice charges, thus generalizing on a graph the notion of magnetic fragmentation in spin ice~\cite{brooks2014magnetic}. 

Let us then call $q_{\text{ext}}=\mu \mathrm{div}[H]$ the ``external charge" (here $\mu$ has dimension of an energy) and assume $\sum_v q_{\text{ext},v}=0$. Then from Eq.~(\ref{QFT5}), integrating over $\phi$ one has
\begin{align}
\langle q\rangle&= -\frac{\mu}{T}\hat W^q  \hat L^{-1}q_{\text{ext}} =-\frac{\mu}{\epsilon}  \hat G q_{\text{ext}}. 
\label{screen}
\end{align}
Thus, an external charge is screened by the screened Green function of the Laplacian. Note that because $\hat L \Psi = - q_{\text{ext}}$ we can also write
\begin{align}
\langle q_v \rangle&= -\frac{\mu}{T} \langle q_v q_{v'}\rangle \Psi_{v'}.  
\label{screen2}
\end{align}
From Eq.~(\ref{screen}) we obtain the two correlation limits
\begin{align}
\langle q_v\rangle & = - \frac{\mu}{\epsilon} \xi^2 \delta_{v  v'} q_{\text{ext}} +O(\xi^{2}) ~~\mathrm{for}~~ \xi\to 0 \nonumber \\
\langle q_v\rangle & = - \frac{\mu}{\epsilon} \left[L^{-1}\right]_{v  v'} q_{\text{ext},v'}+ O(\xi^{-2}) ~~\mathrm{for}~~ \xi \to \infty.
\end{align} 

\subsection{Entropic Screening of Pinned Charges}

The case of a pinned charge  is obtained by summing the partition function only over spin configurations corresponding to a fixed charge $q_{\mathrm{pin}}$ on a  vertex $\bar v$. This corresponds to inserting a $\delta(q_{\bar v} -q_{\mathrm{pin}})$ in the functional integral. We  leave to the reader the simple calculation, whose result is 
\begin{align}
\langle q_v\rangle= \frac{W_{v \bar v}}{W_{\bar v \bar v}} q_{\mathrm{pin}} = \frac{\langle q_v q_{\bar v}\rangle}{\langle  q_{\bar v}^2\rangle}q_{\mathrm{pin}},
\label{screenp}
\end{align}
 and correctly yields $\langle q_{\bar v}\rangle=q_{\mathrm{pin}}$. Note also, from Eq.~(\ref{W2}),
\begin{align}
\langle q_v\rangle & = L_{v \bar v} \frac{q_{\mathrm{pin}}}{z_{\bar v}} +O(T^{-2})  ~~\mathrm{for}~~ T \to \infty.
\end{align} 
The difference between the two screenings should not surprise. In the first case an external field interacts locally [see Eq.~(\ref{Hel})] with the spin ice, inducing local effects that then propagate via the charge-charge correlation, while the second case is due to correlations of the free charge with the pinned one.

\section{Examples} 

\subsection{Star Graph}

A star graph is made of $n$ nodes $v_i$ each connected only a central node $w$. Thus, $N_l=n, N_v=n+1$. The number of ice rule configurations are  $W=\binom{n}{n/2}$ when $n$ is even and $W=2\binom{n}{(n+1)/2}$ when $n$ is odd. This leads to a Pauling entropy per spin $s=\ln(2)$ in the limit of large $n$. The fact that the Pauling entropy coincides with the entropy at infinite temperature is a feature of spin ices with infinite spins per vertex. In particular, to have a finite energy per spin we need to scale the energy coupling as $\epsilon = J /n$. 

 To find the charge correlations, note immediately that by symmetry it must be $\langle q_v \rangle =0$,  $\langle q_w \rangle =0$,  and  $\langle q_v^2 \rangle =1$ for every $v$. The only possible correlations are $\langle q_v^2 \rangle =1$, $\langle q_v q_w \rangle$, $\langle  q_w^2 \rangle$ and $\langle q_v q_{v'} \rangle$. Using tricks such as $\langle q_v q_w \rangle = n^{-1}\sum_{v} \langle q_v q_w \rangle=n^{-1}\langle \sum_{v} q_v q_w \rangle$ and charge neutrality one readily finds
\begin{align}
 &\langle q_v q_w \rangle =- \langle q_w^2 \rangle/n \nonumber \\
 &\langle q_v q_{v'} \rangle = \langle q_w^2 \rangle/[n(n-1)]-1/(n+1), 
 \label{starcor}
 \end{align}
and we need to  find only $\langle q_w^2 \rangle$. Note that $Z=\exp(\beta J /2)Z_w$, where the latter is given by
\begin{align}
Z_w=\sum_{k=0}^n \binom{n}{k}e^{-\frac{J}{2n}(2k-n)^2} \sim\int_{-1}^{1}d\rho e^{-\beta n f(\rho)}
\label{Zw}
\end{align} 
 in the limit of large $n$ [$\rho=(2k-1)/n$, $f(\rho)=J \rho^2/2- T s(\rho)$ with $s(\rho)=\sigma(1/2+\rho/2)$, while $\sigma(x)= -x\ln(x)-(1-x)\ln(1-x)$ is the usual binomial entropy]. In the limit of large $n$ we can perform the quadratic expansion around the minimum of $f$ (which is at $\rho=0$) obtaining $f(\rho)\sim (1/2)(\beta J +1) \rho^2$, and thus $\langle \rho^2 \rangle=n^{-1} (\beta J + 1)^{-1}$. Then from $q_w=n \rho$ and from the Eqs.~(\ref{starcor}) we have
\begin{align}
\langle q_w^2 \rangle =&\frac{n}{1+\beta J} \nonumber \\
\langle q_v q_w \rangle =&\frac{-1}{1+\beta J} \nonumber \\
\langle q_v^2  \rangle =&1 \nonumber \\
\langle q_v q_{v'} \rangle =&   \frac{-1}{n} \frac{\beta J}{1+\beta J} ~~\text{for} ~~v\ne v'.
 \label{starcor2}
 \end{align}

Note that $\langle q_w^2 \rangle \to z_w$ at large $T$ and $\langle q_w^2 \rangle, \langle q_v q_w \rangle \to 0$ at low $T$, as expected. Note also that the Eq.~(\ref{qcorr4}) is valid: from Eq.~(\ref{starcor2}), $\langle q_v q_v' \rangle =-\xi^2 + O(\xi^4)$ and indeed the number of geodesics between $v$ and $v'$ is one. 

What are the correlations deduced from the field theory at high $T$? We leave to the reader the simple task of computing the matrix $\hat W^q$ and verifying that
\begin{align}
W^q_{ww} =&\frac{n^2}{n+(n+1)\beta J} \nonumber \\
W^q_{vw} =&- \frac{n}{n+(n+1)\beta J} \nonumber \\
W^q_{vv}  = &\frac{1+\beta J}{1+\frac{n+1}{n} \beta J + \frac{n+1}{n^2}(\beta J)^2}\nonumber \\
W^q_{vv'} =&  -\frac{\beta J}{n+(n+1) \beta J + \frac{n+1}{n}(\beta J)^2}~~\text{for} ~~v\ne v'.
\label{starcor3}
\end{align}

The equations above correctly reduce to the Eqs.~(\ref{starcor2}) in the limit of large $n$ in which the latter were derived. Therefore, the field theory is exact {\em at any temperature} for the star graph in the thermodynamic limit. That should not surprise. We know that field theories become exact for Ising models of binary spins when each spin interact with each other in the same way, that is when the graph connecting interacting spins is a complete graph, which is the case for the star graph. 

Finally, pinning a charge $Q_p$ on $w$ elicits, from Eq.~(\ref{screenp}) a screening charge $\langle q_v \rangle=-Q_p/n$, which can also be obviously deduced from charge cancellation. More interesting is when the charge is pinned on $v$. Then it must be $Q_p=\pm1$, and the screening is: $\langle q_w \rangle=-Q_p(1+\beta J)^{-1}$, $\langle q_{v'\ne v} \rangle=-(Q_p/n) \beta J /(1+\beta J)$, going to zero in the thermodynamic limit.

\subsection{Complete Graph}

In a complete graph $K_n$ all nodes are connected to all others. We have $N_v=n, N_l=n(n-1)/2$. As already explained, a spin configuration configuration is a digraph called a tournament, and there are $|{\cal P}|=2^{n(n-1)/2}$ tournaments. When $n$ is odd, vertex coordination is even and an ice-rule obeying tournament is called a {\em regular} tournament. Thus $|{\cal I}|$ is the number of regular tournaments with $n$ players, which is still an open problem. There are, however, asymptotic formulas for large $n$. From McKay's formula~\cite{mckay1990asymptotic} we find that the Pauling entropy per spin in the thermodynamic limit is $s=\ln(2)$. As in the star graph is coincides to the entropy at infinite temperature. 

By the symmetry of the problem there are only two charge correlations, $\langle q_v q_{v'} \rangle$ (when $v\ne v'$) and  $\langle q_v^2 \rangle$, and they are related: $\langle q_v q_{v'} \rangle=(n-1)^{-1} \sum_{v'\ne v}\langle q_v q_{v'} \rangle=- \langle q_v^2 \rangle/(n-1)$.

Note that, unlike in a star graph, not all spin interact with other spins. However, its partition function can be factorized in terms corresponding to star graphs in the limit of large $n$. Indeed, consider $K_{n-1}$, then $K_n$ can be former by adding a vertex $v_n$ and $n-1$ edges connecting it to the $v_1 \dots v_{n-1}$ vertices of $K_{n-1}$. Its partition function is therefore
\begin{align}
&Z_n=\sum_{k=0}^n {\bigg [} \binom{n}{k} e^{ -\frac{\beta \epsilon}{2} (2k-n)^2} \times \nonumber \\
&\sum_{\{S \} \in K_{n-1}}e^{ -\frac{\beta \epsilon}{2}\left[\sum_{j=1}^k (q_j-1)^2 +\sum_{j=k+1}^{n-1} (q_j+1)^2\right]}  {\bigg ]} \nonumber \\
&~~=Z_{n-1}\sum_{k=0}^n \binom{n}{k} e^{ -\frac{\beta \epsilon}{2} (2k-n)^2+j_k},
\end{align}
where $j_k=\ln \langle e^{ {2\beta \epsilon}\sum_{j=1}^k q_j}\rangle_{n-1}$, and the suffix indicates that the average is performed on the $K_{n-1}$ graph. Notice than, again, we must assume $\epsilon=J/n$ in order to have an extensive energy. Then, with the same transformation as for the star graph, $\beta \epsilon (2k-n)^2=n \beta J \rho$ scales as $n$, whereas $j_k$ is subextensive in $n$ and can be neglected. Thus, for large $n$, $Z_n=Z_{n-1}Z_w$, with $Z_w$ given by Eq.~(\ref{Zw}). We obtain therefore, as before
\begin{align}
\langle q_v^2 \rangle =&\frac{n}{1+\beta J}, 
 \label{Kcorr}
 \end{align}
and from $\langle q_v q_{v'} \rangle=-\langle q_v^2 \rangle/(n-1)$  we have
\begin{align}
\langle q_v q_{v'} \rangle =&  - \frac{1}{1+\beta J}.
\label{Kcorr2}
\end{align}

Note in particular that while the entropy is extensive in the number of spins, energy scales linearly in the number of vertices. Indeed $E=\langle \epsilon \sum_v q_v^2\rangle/2=(n/2)JT/(T+J)$: the energy per vertex is finite and non zero (except for $T=0$) while the energy per spin is always zero. 

From Eq.~(\ref{screenp}) we see that by pinning a charge $Q_p$ in a vertex, the charge elicited by entropic screening in all other vertices is 
 %
$\langle q \rangle =-{Q_p}/{n}$,
%
which is obvious since there must be charge cancellation. The elicited charge is zero in the thermodynamic limit if $Q_p$ is finite, and non-zero if $Q_p$ instead scales linearly in $n$.

Finally, the adiacency matrix for the complete graph $K_n$ is $\hat A= \hat J -\hat 1$, and thus the Laplacian is $\hat L = (n-1) \hat 1- \hat J$, where $\hat J$ is the matrix of ones. From that it is immediate to find $\hat W^q$ as 
\begin{align}
W^q_{vv}  = &\frac{n-1}{1+ \beta J} \nonumber \\
W^q_{vv'} =& -\frac{1}{1+ \beta J}~~\text{for} ~~v\ne v'.
\label{Kcorr3}
\end{align}
which corresponds to the correlations of Eqs.~(\ref{Kcorr}, \ref{Kcorr2}) for $n$ large. Again, we see that the field theory is exact at any temperature. 

\subsection{Spin Ice on a Lattice}

We will treat elsewhere the cases of spin ice on a lattice in full. Here we report general considerations that are common to spin ices that 
 can be properly embedded in a linear space of dimension $d$, such that in the thermodynamic limit it can be homogenized to a continuum ($v\to \vec x$), and distances are measured in units of the lattice constant such that Eq.~(\ref{QFT6}) becomes 
\begin{align}
{\cal F}_2[q]&=\frac{ \epsilon}{2} \int q(x)^2\text{d}^d \!x + T \int \text{d}^d \!x \text{d}^d \!y ~  q(x) V^e (x-y) q(y).
\label{QFT6}
\end{align}
Because the Laplacian operator of the graph coincides with that of the linear space, its eigenvectors are plane waves and at small momentum (or large distances) and the eigenvalues are  $\gamma(k)^2 \simeq\vec{k}^2$ (we measure space in units of the lattice constant). Then we have for the entropic interaction at large distances
\begin{equation}
V^e(x)=\begin{cases}
              -\frac{T}{2} |x| ~~~~~~~~~~~d=1\\
              -2 \pi T\ln|x|~~~\!~~d=2\\
	      \frac{T}{4 \pi \xi^{4}x} ~~~~~~~~~~~~d=3,\\            \end{cases}
\label{coulomb}	  
\end{equation}
which are all Coulomb potentials in their proper dimension, as already anticipated in section III. The third line coincides with the numerically verified entropic interaction for pyrochlore spin ice~\cite{castelnovo2011debye,chern2014realizing}.

From Eq.~(\ref{qcorr1}) we obtain
\begin{equation}
\langle |q(k)|^2\rangle = \frac{k^2}{1+\xi^2 k^2}=\xi^{-2} -\xi^{-4}\frac{1}{k^2+\xi^{-2}},
\label{corrSI}
\end{equation}
from which we have, via Fourier transform,
\begin{equation}
\langle q(x) q(0)\rangle=\begin{cases}
              -\frac{1}{2\xi^{3}} \exp\left({-|x|/\xi}\right)~~~~~~~~~~~~d=1\\
              -\frac{1}{2 \pi \xi^{4}} K_0\left(|x|/\xi\right)~~~~~~~~~~~~~~d=2\\
	      -\frac{1}{4 \pi \xi^{4}x} \exp\left({-|x|/\xi}\right)~~~~~~~~~d=3,\\            \end{cases}
\label{corrSP}
\end{equation}
valid at large distances. 
The first line corresponds to the exponential form of the screening in a 1D Ising system, where domain walls correspond to charges. We note that $\xi$ thus known exactly as $\xi_{\text{1D-Ising}}=-1/\ln \tanh (\beta\epsilon)$ and does not follow the $\sim \sqrt{\epsilon/T}$ at high $T$. Though one can verify that at low $T$ we have $\xi_{\text{1D-Ising}}\sim1/\langle |q| \rangle$ which of course is different from $\xi^2 \sim1/\langle q^2 \rangle$. In general, Eq.~(\ref{xi0}) underestimates the correlation in the one-dimensional case because one-dimensional spin ice has an ordered ice manifold. As we show in future work, it enjoys a unique position in such regard. 

The second line in Eq.~(\ref{corrSP}) was recently experimentally verified in square spin ice realized on a quantum annealer by pinning a charge and relaxing the system~\cite{king2020quantum}.

\section{Conclusions}

We have proposed the concept of spin ice on a general graph, and related it to well known graph-theoretical concepts of balance and Eulerian paths. We have then developed a field theoretical framework for its excitations, monopoles or charges. We have obtained a series of results that are independent of geometry. The partition function of general Graph Spin Ice can be reformulated exactly as a functional integral over the charge distribution and its entropic field, the latter subsuming the effect of the underlying spin ensemble on emergent charges [Eqs~(\ref{Z2})-(\ref{QFT3})]. The high $T$ behavior  is described by a quadratic free energy of the average charges and contains informations on the graph via the graph Laplacian [Eq.~(\ref{QFT6})].

In absence of charge interaction and external fields and in the limit of high $T$, the entropic interaction among charges corresponds to the Green operator of the graph Laplacian. Thus, correlations correspond to the {\it screened} Green operator of the graph [Eqs~(\ref{qcorr5})-(\ref{qcorr1})]. The correlation length is $\xi^2=\epsilon/T$ at high $T$ and $\xi^2=1/\langle q^2 \rangle$ at low $T$, results already appreciated in various special spin ice systems~\cite{garanin1999classical,fennell2009magnetic}.

Beside generalizing condensed matter notions to graphs where they can provide intuitive insight or inspiration for broader problems in complex networks, we have shown that many of the properties of spin ice systems follow directly from the graph structure, which is essentially topological.

\section{Acknowledgements}
We thank  Beatrice Nisoli for proofreading.
This work was carried out under the auspices of the U.S.
DoE through the Los Alamos National
Laboratory, operated by
Triad National Security, LLC
(Contract No. 892333218NCA000001).

\bibliography{library2.bib}{}

\end{document}